\documentstyle[11pt]{article}
\newcommand{\IcA}{{\Im m\cal A}}
\newcommand{\RcA}{{\Re e\cal A}}
\newcommand{\Frac}[2]{\frac{\displaystyle #1}{\displaystyle #2}}
\newcommand{\kpiggp}{$K^+ \rightarrow \pi^+ \gamma \gamma $ }
\newcommand{\kpiggtot}{$K \rightarrow \pi \gamma \gamma $ }

\newcommand{\kpiggn}{$K_L \rightarrow \pi^0 \gamma \gamma $ }

\newcommand{\kgll}{$K_L \rightarrow  \gamma \ell^+ \ell^- $ }

\newcommand{\kggs}{$K_L \rightarrow  \gamma \gamma^* $ }
\newcommand{\klmm}{$K_L \rightarrow  \mu^+ \mu^- $ }
\newcommand{\opc}{${\cal O}(p^4)$ }
\newcommand{\ops}{${\cal O}(p^6)$ }
\begin{document}
\begin{flushright}
INFNNA-IV-97/46\\
DSFNA-IV-97/46\\
{\tt hep-ph/9709314}
\end{flushright}
\vspace*{1cm}
\centerline{\Large\bf Radiative and rare kaon decays~: an update 
\footnote{To appear in the Proceedings of the XVI International Workshop
on Weak Interactions and Neutrinos (WIN'97) Capri, Italy, June 22-28,
1997.}}
\bigskip
\vspace*{1.5cm}
\centerline{Giancarlo D'Ambrosio ${}^a$}
\bigskip
\centerline{${}^a$INFN, Sezione di Napoli, Dip. di Scienze Fisiche, Univ. 
            di Napoli, I-80125 Napoli, Italy.}
\vspace*{1cm}
\centerline{\bf Abstract}
\vspace*{0.3cm} 
\noindent      
We review some new developments in the theoretical description of the
processes \kpiggtot, \kgll, $K_L \rightarrow \pi^0 e^+ e^-$ and \klmm.
\vspace*{1cm}

\section{Introduction}

Radiative non--leptonic kaon decays may play a crucial role in our 
understanding of fundamental questions: 
the validity of the Standard Model (SM) (fixing  the values of the 
CKM parameters),
the origin of CP violation  
 and as a $\chi$PT (Chiral Perturbation Theory) test
(see \cite{PICH97,DE95,DAGI96} and references therein). 
\par
The \kggs form factor (together with $K_L \rightarrow \gamma^* \gamma^*$)
is an important ingredient  to evaluate properly the dispersive
contribution to the real part of the amplitude for the decay
$K_L \rightarrow \gamma^* \gamma^* \rightarrow \mu^+ \mu^-$. Since this
real part receives also short distance contributions proportional
to the CKM matrix element $V_{td}$ \cite{NEB77} and the absorptive 
amplitude is found to saturate the experimental result, a strong 
cancellation in the real part between short and long distance 
contributions or the addition of two very small amplitudes is expected. 

\par
The importance of \kpiggn goes further the interest of the process in 
itself because its role as a CP--conserving two--photon discontinuity
amplitude to $K_L \rightarrow \pi^{0} e^+ e^-$ in possible 
competition with the CP--violating contributions \cite{DE95,DAGI96,DG95}
which are predicted to be of $O(10^{-12})$. The size 
of the  CP--conserving contribution depends on an helicity
amplitude  for  \kpiggn which appears  only at next-to-leading order
and this contribution can be controlled both theoretical and
experimentally, as we shall see.  
\par
The interplay between experimental results and phenomenology indicates that
in both these decays ($K_L \rightarrow \pi^0 \gamma \gamma$
and $K_L \rightarrow \gamma \gamma^*$) there is an important  vector
meson exchange contribution.
Actually the experimental determination of the slope in \kggs is only
based in the BMS model \cite{BMS83}, 
thus we propose an alternative analysis less model
dependent. 
We realize also that the common problematic point between \kpiggtot and 
\kggs is the model dependence  of the weak 
$VP\gamma$ vertex. Thus after constructing the most general leading $\chi$PT
lagrangian for the weak $VP\gamma$ vertex for the processes under
consideration  we propose a Factorization Model 
in the Vector couplings (FMV).

\section{\kpiggtot and \kggs amplitudes}
\hspace*{0.1cm}
The general amplitude for 
$K_L(p) \rightarrow \pi^0 \gamma(q_1) \gamma(q_2)$
can be written in terms of two independent Lorentz and gauge invariant 
amplitudes~: $A(z,y)$ and $B(z,y)$, where 
$y  = p \cdot (q_1 - q_2)/m_K^2 $ and  
$ z \, = \, (q_1 + q_2)^2/m_K^2 $.
Then the double differential rate  is given by 
\begin{equation}
\Frac{\partial^2 \Gamma}{\partial y \, \partial z}   = 
\, \Frac{m_K}{2^9 \pi^3} [ \, z^2 \, | \, A \, + \, B \, |^2 \,
 + \,  \left( y^2  -  \Frac{\lambda (1,r_{\pi}^2,z)}{4} \right)^2
 \, | \, B \, |^2 \, ]~, 
\label{eq:doudif} 
\end{equation}
where $\lambda (a,b,c)$ is the usual kinematical function and 
$r_{\pi} =  m_{\pi}/m_K $. Thus in the 
region of small $z$ (collinear photons) the $B$ amplitude is dominant
and can be determined separately from the $A$ amplitude. This feature
is important in order to evaluate the CP conserving contribution
$K_L \rightarrow \pi^0 \gamma \gamma \rightarrow \pi^0 e^+ e^-$. Both
on--shell and off--shell two--photon intermediate states generate, through
the $A$ amplitude, a contribution to $K_L \rightarrow \pi^0 e^+ e^-$
that is helicity suppressed \cite{EPR88}. Instead
the $B$--type amplitude, though appearing only at \ops, generates 
 a relevant unsuppressed 
contribution to $K_L \rightarrow \pi^0 e^+ e^-$ through the
on--shell photons \cite{DG95}, due to the different 
helicity structure.
\par
If CP is conserved the decay $K_L \rightarrow \gamma (q_1, \epsilon_1) 
\gamma^* (q_2, \epsilon_2)$  is given by an only amplitude $A_{\gamma 
\gamma^*}(q_2^2)$ that can be expressed as 
$
A_{\gamma \gamma^*} (q_2^2)   =     A_{\gamma \gamma}^{exp} f(x)  $,
where $A_{\gamma \gamma}^{exp}$ is the experimental amplitude 
$A(K_L \rightarrow \gamma \gamma)$ and $x=q_2^2/m_K^2$. The form factor
$f(x)$ is properly normalized to $f(0)=1$ 
and the slope $b$ of $f(x)$ is defined as 
$ f(x)  =  1   +   b x  $. 
Traditionally  experiments \cite{PDG} do not measure directly the slope but 
they input the  full form factor suggested by the BMS model
\cite{BMS83}, however we think that is more appropriate to
measure directly the slope and we estimate \cite{DPN}, 
$
b_{exp} \, = \, 0.81 \pm 0.18 \,
$.
The slope $b$ gets two different contributions: the first one ($b_V$) 
comes from the strong vector interchange with the weak transition
in the $K_L$ leg, the second comes from a direct weak transition
$K_L \rightarrow V \gamma$ ($b_D$). Then $b \, = \, b_V \, + \, b_D $.
While the first term is model independent, the direct contribution
$b_D$ requires a modelization due to our ignorance of the weak couplings
involving vector mesons.
BMS  model \cite{BMS83}  suggests that the
structure of the weak $VP\gamma$ vertex is dominated by a weak 
vector--vector transition. 
\par
The leading finite \opc amplitudes of \kpiggn were evaluated some time ago
\cite{DE87}, generating only the $A$--type amplitude in Eq.~(\ref{eq:doudif}).
The observed branching ratio for \kpiggn is $(1.7 \pm 0.3) \times 10^{-6}$ 
\cite{PDG}
which is  about 3 times  larger than the \opc prediction \cite{DE87,CD93}.
However the \opc spectrum of the diphoton invariant mass nearly agrees with
the experiment, in particular no events for small $m_{\gamma \gamma}$ are
observed, implying a small $B$--type amplitude. Thus \ops corrections have
to be important. Though no complete calculation is available, the 
supposedly larger contributions have been performed~: \ops unitarity 
corrections  
\cite{CD93,CE93,KH94} enhance the \opc branching ratio by $30 \%$,
and generate a $B$--type amplitude.
\par
One can
parameterize the \ops vector meson exchange contributions
to $K_L\rightarrow \pi^0 \gamma \gamma$,  by an
effective vector coupling $a_V$ \cite{EP90} :
\begin{eqnarray}
A & = & {G_8 M_K^2\alpha_{em}\over \pi } a_V(3-z+r_{\pi}^2) ~, \nonumber \\
B & = & -{2G_8 M_K^2\alpha_{em}\over \pi } a_V~,\label{eq:abvec}
\end{eqnarray}
where $G_8$ is the effective coupling of the leading octet weak chiral 
lagrangian and is fixed by $K \rightarrow \pi \pi$.
Analogously to the \kggs case there are two sources for $a_V$~:
i) strong vector resonance exchange with an external weak transition 
($a_V^{ext}$), and
ii) direct vector resonance exchange between a weak and a strong 
$VP\gamma$ vertices ($a_V^{dir}$). Then
$
a_V \, = \, a_V^{ext} \, + \, a_V^{dir} 
$.
The first one is model independent and gives $a_V^{ext} \simeq 0.32$ 
\cite{EP90}, while the direct contribution depends strongly on the model
for the weak $VP\gamma$ vertex.
Cohen , Ecker and Pich noticed \cite{CE93} that one could, 
simultaneously, obtain the experimental spectrum and width of 
\kpiggn with $a_V \simeq -0.9$. The comparison of this result with 
the value of $a_V^{ext}$ shows the relevance of the direct contributions.
\par
The question of the relevant \ops contributions relative
to the leading \opc result for \kpiggp \cite{EPR88} can also be studied.
\ops unitarity corrections \cite{DP96}
generate a $B$--type amplitude
and increase the rate by a  $30$--$40 \%$ while, differently from \kpiggn,
vector meson exchange is negligible in \kpiggp \cite{DPN}.
 BNL--787 has got, for the first time events in this channel
\cite{TAK96} confirming the relevance of the unitarity corrections.

\section{Factorization Model in the Vector Couplings (FMV)}

The general effective weak coupling
$VP\gamma$  contributing to \ops \kpiggtot and \kggs processes is 
\begin{equation}
{\cal L}_W(VP\gamma) \; = \; G_8 \, F_{\pi}^2 
\, \varepsilon_{\mu \nu \alpha \beta} \,
\sum_{i=1}^{5} \, \kappa_i \, \langle \, V^{\mu}
\,   T_i^{\nu \alpha \beta} \rangle \; ,
\label{eq:mostg}
\end{equation}
where $T_i^{\nu \alpha \beta}$ $i=1,..,5$ are all the possible 
relevant $P\gamma$ structures \cite{DPN}, the $\kappa_i$ are the dimensionless
coupling constants to be determined from phenomenology or theoretical
models,  $F_{\pi} \simeq 93 \, \mbox{MeV}$ is the pion decay constant and the
brackets in Eq.~(\ref{eq:mostg}) stand for a trace in the flavour space.
\par
Motivated by the $1/N_c$ expansion the Factorization Model (FM) \cite{PI91} 
assumes that the dominant contribution to the four--quark operators
of the $\Delta S=1$ Hamiltonian comes from a factorization {\em current 
$\times$ current} of them. This assumption is implemented with a 
bosonization of the left--handed quark currents in $\chi$PT as given by
\begin{equation}
\overline{q_{jL}} \gamma^{\mu} q_{iL} \; \longleftrightarrow
\; \Frac{\delta \, S [U,\ell,r,s,p]}{\delta \, \ell_{\mu, ji}}~,
\label{eq:boscur}
\end{equation}
where $i,j$ are flavour indices, and $S$ is the 
low--energy strong effective action of QCD in terms of the Goldstone
bosons realization $U$ and the external fields $\ell, r, s, p$.
\par
The general form of the lagrangian  is
\begin{equation}
{\cal L}_{FM} \; = \; 4 \, k_F \, G_8 \, \langle \, \lambda \, 
\Frac{\delta S}{\delta \ell_{\mu}} \, \Frac{\delta S}{\delta \ell^{\mu}} \, 
\rangle \; + \; h.c. \; \; \; , 
\label{eq:fmgenera}
\end{equation}
where $\lambda \, \equiv \, \frac{1}{2} \, (\lambda_6 \, - \, i \lambda_7)$.
The FM  gives a full prediction but for a fudge factor
$k_F \sim {\cal O}(1)$. 
\par
The procedure used in the FM to study the resonance exchange contribution
to a specified process involves first the construction of a 
resonance exchange generated strong Lagrangian, in terms of Goldstone bosons
and external fields, from which to evaluate the left--handed currents.
For example, in \kpiggtot one starts from the strong $VP\gamma$ vertex
and integrates out the vector mesons between two of those vertices 
giving a strong Lagrangian for $P P \gamma\gamma$ ($P$ is short for
pseudoscalar meson) from which to evaluate the left--handed current.
This method of implementing the FM imposes as a constraint
the strong effective $P P \gamma \gamma$ vertex and therefore
it can overlook parts of the chiral structure of the weak vertex.
In the FM model the dynamics is hidden in the fudge factor $k_F$. 
Alternatively one could try to identify the different vector meson
exchange contributions and then estimate the relative weak
couplings.
\par
We propose precisely this scheme. Instead of getting the strong 
lagrangian generated
by vector meson exchange we apply the factorization procedure to the
construction of the weak $VP\gamma$ vertex and we integrate out the
vector mesons afterwards. The interesting advantage of our approach is that, 
as we have seen in the processes we are interested in, it allows us to 
identify \underline{new contributions} to the left--handed currents and 
therefore to the chiral structure of the weak amplitudes. This we call
the Factorization Model in the Vector couplings (FMV).
\par
Hence we obtain the $\kappa_i$ in Eq.~(\ref{eq:mostg}) in this framework
and  we can give a prediction from the FMV model to 
both $a_V^{dir}$ in \kpiggn and the slope $b_D$ of \kggs \cite{DPN}. We get
$
a_{V}^{dir} \, \left|_{FMV}  \, \simeq \, -0.95 \right.
$ 
and
$
b_D^{octet} \; \left|_{FMV} \, \simeq \, 0.51 \right.
$.
Our final predictions (see Ref. \cite{DPN} for a thorough discussion)
are~:
\begin{equation}
a_V \, \simeq  \, -0.72 \; \; \; \; \; , \; \; \; \; \; 
b \,  \simeq  \, 0.8 - 0.9~, 
\label{eq:finres} 
\end{equation}
in good agreement with  phenomenology. Thus we can  predict 
the CP conserving 
contribution: $0.1<B(K_L \rightarrow \pi^0 e^+ e^-)\cdot 10^{12}<3.6 $ for
  $-1.0<a_V<-0.4$
respectively \cite{DPN}.

\section{Dispersive two--photon $B(K_L \rightarrow \mu^+ \mu^-)$}

To fully exploit the potential of \klmm in probing short--distance 
dynamics it is necessary to have a reliable control on its long--distance
amplitude. However the dispersive contribution generated by the 
two--photon intermediate state cannot be calculated in a model independent
way and it is subject to various uncertainties \cite{VS}. The branching
ratio can be generally decomposed as 
$
B(K_L \rightarrow \mu^+ \mu^-) \, = \, |\RcA|^2 \, + \, |\IcA|^2
$,
and the dispersive contribution can be rewritten as 
$\RcA \, = \, \RcA_{long} \, + \, \RcA_{short}$.
The recent measurement of $B(K_L \rightarrow \mu^+ \mu^-)$ 
\cite{BNL} is almost saturated by the absorptive amplitude leaving
a very small room for the dispersive contribution~:
$
| \RcA_{exp} |^2 \, = \, (-1.0 \pm 3.7) \times 10^{-10}
$ or
$| \RcA_{exp} |^2 \, < \, 5.6 \times 10^{-10} $ at $90 \%$ C.L.
\par
Within the Standard Model the NLO short-distance amplitude \cite{BB} gives
the possibility to extract a lower bound on $\overline{\rho}$ 
($\overline{\rho} = \rho ( 1 - \lambda /2)$ and $\rho$ and $\lambda$
the usual Wolfenstein parameters) as
\begin{equation}
\overline{\rho}  >  1.2 - max  \left\{ 
\Frac{|\RcA_{exp}| + |\RcA_{long}|}{3 \times 10^{-5}} \, 
\left[ \Frac{\overline{m_t}(m_t)}{170~\mbox{GeV}} \right]^{-1.55}
\left[ \Frac{|V_{cb}|}{0.040} \right]^{-2} \right\}~.
\label{eq:bound} 
\end{equation}
In order to saturate this lower bound we propose \cite{DIP} a low
energy parameterization of the $K_L \rightarrow \gamma^* \gamma^*$
form factor that include the poles of the lowest vector meson 
resonances with arbitrary residues
\begin{equation}
f(q_1^2,q_2^2)  =  1 + \, \alpha \left( \Frac{q_1^2}{q_1^2 - m_V^2} +
\Frac{q_2^2}{q_2^2 - m_V^2} \right) \,  + \, \beta \,  
\Frac{q_1^2 q_2^2}{(q_1^2 - m_V^2) (q_2^2 - m_V^2)}~.
\label{eq:fq1q2} 
\end{equation}
The parameters $\alpha$ and $\beta$, expected to be ${\cal O}(1)$
by naive dimensional chiral power counting, are in principle directly
accessible by experiment in \kgll and $K_L \rightarrow 
e^+ e^- \mu^+ \mu^-$. Up to now there is no information on $\beta$ and
we have the phenomenological result $\alpha = -1.63 \pm 0.22$. The 
form factor defined in Eq.~(\ref{eq:fq1q2}) goes as $1 + 2 \alpha + 
\beta$ for $q_i^2 \gg m_V^2$
and one has to introduce an ultraviolet cutoff $\Lambda$.
However, in this region the  perturbative QCD  calculation
of the two-photon contribution gives a small result. In particular we find
$|1+2\alpha+\beta|\ln (\Lambda/M_V) <  0.4$,
showing a mild behaviour of the form factor at large $q^2$. The FMV
model described in the previous Section gives $(1+2\alpha+\beta)_{FMV} 
\simeq - 0.01$ in good agreement with the perturbative result result.
\par
Using $\alpha_{exp}$ and the QCD constraint we predict \cite{DIP}
\begin{equation}
| \RcA_{long} | < 2.9 \times 10^{-5} \; \; \; \; (90 \% C.L.)
\label{eq:rcabo}
\end{equation}
and
\begin{equation}
\overline{\rho} > -0.38~, \; \; \; \; \mbox{or} \; \; \; \; 
\rho > -0.42 \; \; \; (90 \% C.L.)~.
\label{eq:rhobo}
\end{equation}
These bounds could be very much improved if the $\alpha$ and $\beta$
parameters were measured with good precision and a more stringent
bound on $|\RcA_{exp}|$ is established.
\vspace*{-0.2cm} 
\section{Acknowledgement}
\vspace*{-0.2cm} 
I wish to 
thank G.  Isidori and J. Portol\'es for the very fruitful collaborations
and discussions.

\end{document}